\documentclass[aps,twocolumn,prb,superscriptaddress,showpacs,eqsecnum]{revtex4}
\usepackage{amsmath,amssymb,latexsym}
\usepackage[dvips]{graphicx}

\begin{document}

\title{Excitonic effects in optical absorption and
electron-energy loss spectra of hexagonal boron nitride}

\author{Ludger Wirtz}
\affiliation{Institute for Electronics, Microelectronics, and Nanotechnology
(IEMN), CNRS-UMR 8520, B.P. 60069, 59652 Villeneuve d'Ascq Cedex, France
}

\author{Andrea Marini}
\affiliation{Istituto Nazionale per la Fisica della Materia e 
Dipartimento di Fisica dell'Universit\'a di Roma ``Tor Vergata'',
Via della Ricerca Scientifica, I-00133 Roma, Italy}

\author{Myrta Gr\"uning}
\affiliation{Donostia International Physics
Center (DIPC), 20018 Donostia-San Sebasti\'an, Spain}

\author{Angel Rubio}
\affiliation{Donostia International Physics
Center (DIPC), 20018 Donostia-San Sebasti\'an, Spain}
\affiliation{Department of Material Physics, UPV/EHU and Centro Mixto CSIC-UPV,
20018 San Sebasti\'an, Spain}
\affiliation{Institut f\"ur Theoretische Physik, Freie Universit\"at Berlin,
Arnimallee 14, D-14195 Berlin, Germany}

\date{\today}

\begin{abstract}
A new interpretation of the optical and energy-loss
spectra of hexagonal boron nitride is provided 
based on first-principle calculations.  We show that both
spectra cannot be explained by independent-particle transitions
but are strongly dominated by excitonic effects.
The lowest direct and indirect gaps are much larger than previously 
reported. The direct gap amounts to 6.8 eV. 
The first absorption peak at 6.1 eV 
is due to an exciton with a binding energy of 0.7 eV.
We show that this strongly bound Frenkel exciton 
is also responsible for the low frequency shoulder of the $\pi$ plasmon
in the energy-loss function. Implications for nanotube studies are discussed.
\end{abstract}

\pacs{78.20.-e, 71.35.Cc, 71.45.Gm}

\maketitle

Hexagonal boron nitride (hBN) is isoelectronic to graphite
and has a similar layered structure.
Recently, the electronic structure and the optical properties
of hBN have regained interest due to the discovery of BN-nanotubes 
\cite{rub94,cho95} which can be considered as cylinders formed by rolling 
a single sheet of hBN onto itself.
While graphite is a semi-metal with a linear crossing of the $\pi$
and $\pi^*$ bands at the K-point, in hBN the different electronegativities 
of boron and nitrogen lift the degeneracy at the K-point and lead to a 
large gap. It was predicted \cite{blasebn} that hBN has an 
indirect quasi-particle gap of 5.4 eV and a minimum direct quasi-particle
gap of 6.2 eV. Measurements of the optical absorption coefficient
\cite{zung76,Hoffman,tarrio,lauret,arenal} yield unanimously an
absorption peak with a maximum between 6.1 and 6.2 eV. 
The spectra in these papers are interpreted in terms of independent 
electron transitions from the valence to the conduction band.
The onset of the absorption peak
(varying in the different experiments between 5.2 and 5.5 eV) is 
consequently considered to be a measure of the band gap. 
However, calculations \cite{rpa-bn} at the 
Random Phase Approximation (RPA) level, i.e., in the picture of independent
electron excitations, yield spectra whose shape is not in good
agreement with the experimental spectra. Furthermore, in a recent
luminescence experiment \cite{watanabe}, evidence for a direct bandgap 
at 5.97 eV
and a series of bound excitons with a maximum binding energy of
0.149 eV was found.
In this communication, we reinterpret the different experimental results
and demonstrate the importance of excitonic effects for a proper
understanding of both absorption and electron-loss spectra.

\begin{figure*}
  \includegraphics*[width=1.0\textwidth]{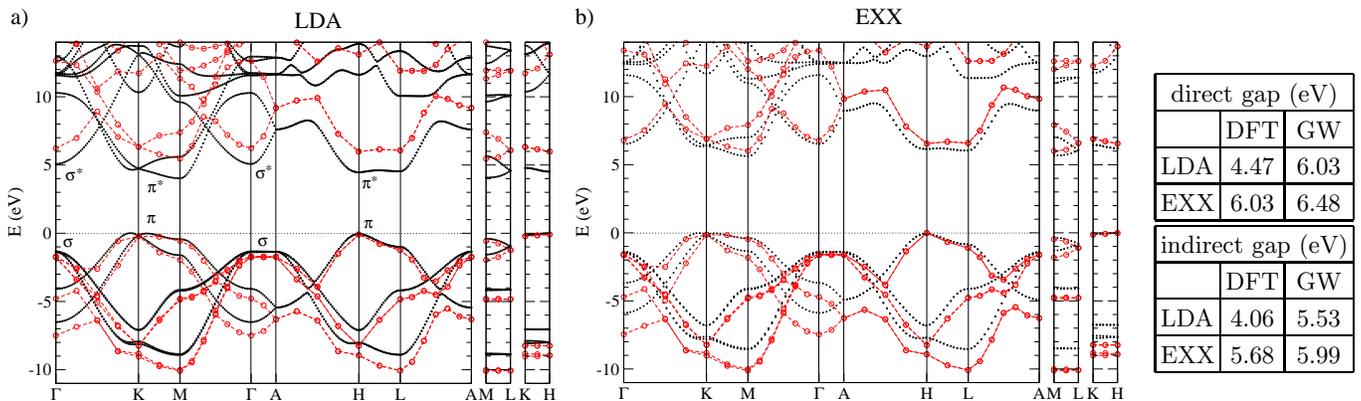}
  \caption{Bandstructure of hBN: a) LDA (dotted lines), LDA-GW (circles);
b) EXX (dotted lines), EXX-GW (circles). Right panel: summary of the direct and indirect gap for hBN in the different approximations of panels a) and b).
}
\label{bandstruc}
\end{figure*}

For the interpretation of the spectra, we calculate the
real and imaginary parts, $\epsilon_1(E)$ and $\epsilon_2(E)$,
of the energy-dependent macroscopic dielectric function on
two levels of approximation: i.) using RPA,
ii) on the GW+BS level, i.e., including quasi-particle corrections
on the level of the GW-method and calculating excitonic effects
through the Bethe-Salpeter (BS) equation \cite{rmp}.
In both cases, we start with a calculation of the wavefunctions
using density-functional theory (DFT). With the code
{\tt ABINIT} \cite{abinit} we calculate \cite{details}
the valence-band wavefunctions and
a large number of conduction-band states, $\psi_{nk}$,
with a band-index $n$ and a sufficiently fine sampling
of the crystal momentum ${\bf k}$.
In this work we use two descriptions of exchange-correlation effects:
the customary local density approximation (LDA) and the exact-exchange
(EXX) potential. To obtain the EXX potential we followed 
Ref.~\onlinecite{stadele}.

On the level of the RPA (which corresponds to the use of Fermi's
golden rule for $\epsilon_2(E)$), we compute the optical spectra
in terms of the dipole-matrix averaged joint-density-of-states.
However, ground-state DFT provides only a qualitative and, often
incorrect, description of the excited states. The reason is
that many-body effects beyond the independent
particle approximation are neglected. 
But in systems with a poor electronic screening,
the electron-hole attraction can be far from negligible.
Good agreement with experiments is restored
by using the self-energy approach of many body perturbation theory \cite{rmp}.
Starting from the DFT energies and wavefunctions, we calculate the 
quasi-particle energies
(``true'' single-particle excitation energies), $E_{nk}$,
by solving the Dyson equation:
\begin{equation}
\left[ -\frac{\nabla^2}{2}
+ V_{ext} + V_{Hartree} +
\Sigma(E^{qp}_{n{\bf k}}) \right]
\psi^{qp}_{n{\bf k}} = E^{qp}_{n{\bf k}} \psi^{qp}_{n{\bf k}}.
\end{equation}
The self-energy $\Sigma$  is approximated as the energy convolution 
of the one-particle Green's function $G$ and the (RPA) dynamically
screened Coulomb interaction $W$, ($\Sigma=iGW$).
We perform a ``semi-self consistent''
(GW$_0$) calculation by updating the quasi-particle energies in $G$ 
(but not in $W$) until the resulting quasi-particle energies are converged.
This procedure yields a quasi-particle bandgap for hBN that is about
0.3 eV higher than the one obtained on the non-self consistent (G$_0$W$_0$) 
level. In order to asses how important are the initial eigenfunctions in
this procedure, we performed calculations starting from LDA or EXX states.

Electron-hole attraction (excitonic effects) is included
by solving the Bethe-Salpeter equation\cite{rmp}
\begin{equation}
(E^{qp}_{c{\bf k}}-E^{qp}_{v{\bf k}}) A^S_{vc{\bf k}} +
\Sigma_{{\bf k'}v'c'} \left\langle
vc{\bf k}|K_{eh}|v'c'{\bf k'}\right\rangle A^S_{v'c'{\bf k'}} =
\Omega^S A^S_{vc{\bf k}}.
\end{equation}
$A^S_{v{\bf k}}$ are the excitonic eigenstate projections onto the electron-hole
basis. The interaction kernel $K_{eh}$ ``mixes'' different
electron transitions from valence band states $v,v'$ to conduction
band states $c,c'$ leading to modified transition energies
$\Omega^s$. The overall effect in the spectra
is a redistribution of oscillator strength as well as the appearance of
new-states within the bandgap (bound excitons).
The GW+BS calculations are performed with the 
code {\tt SELF} \cite{self}.

In Fig.~\ref{bandstruc}, we present the bandstructure of hBN
obtained using the LDA and EXX approximations and compare with
the GW-calculation. On the LDA and GW levels,
we reproduce quite accurately the results of Ref.~\onlinecite{blasebn}.
The GW-correction increases the direct bandgap (at the M-point) by 1.56 eV.
Fig.~\ref{bandstruc} also includes the
values for the direct and indirect gaps~\cite{gwarnaud}.
The bare EXX bandstructure resembles very closely the LDA GW-bandstructure
The GW-correction
on top of EXX wave-functions yields an additional increase of the
direct bandgap of 0.45 eV.  However it still underestimates the
``true'' direct bandgap that must be close to 6.8~eV in order
to describe the measured optical spectra (see below).

\begin{figure}
  \includegraphics*[width=.5\textwidth]{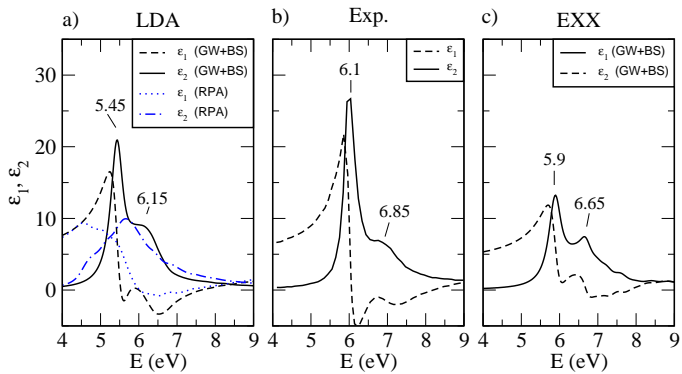}
  \caption{Real ($\epsilon_1$) and imaginary ($\epsilon_2$) parts
of the dielectric function of hBN  calculated using LDA (a)
or EXX (c). The experimental data from Ref.~\onlinecite{tarrio} is shown in (b).
For the LDA results in a) we add the comparison with the results obtained
at the RPA and full GW+BS levels.
The calculations include a Lorentzian broadening
of 0.2 eV (full-width at half-maximum).
The light-polarization is parallel to the BN layers.}
\label{roughspecs}
\end{figure}

Our results  for the energy-dependent dielectric function of hBN
(light polarization parallel to the layers) are shown in
Fig.~\ref{roughspecs} and compared with the
experimental data of Ref.~\onlinecite{tarrio}.
The experimental $\epsilon_1$ and $\epsilon_2$ were extracted from
the electron-energy loss function through a Kramers-Kronig
analysis\cite{tarrio}. 
Since the experimental broadening is about 0.2 eV,
we use the same value in our calculations.
In Fig.~\ref{roughspecs} a) we present calculations based on
LDA-wavefunctions. 
The dash-dotted line shows the RPA absorption spectrum (in
agreement with earlier RPA calculations \cite{rpa-bn}).
The broad peak with a maximum at 5.6 eV is entirely due to the continuum of
inter-band transitions between the $\pi$ and $\pi^*$ bands (see 
Fig.~\ref{bandstruc}).
The calculated GW+BS absorption spectrum displays a double peak structure with
the main peak at 5.45 eV and a second peak at 6.15 eV. 
The nature of the spectrum is entirely different from the RPA
spectrum: the first peak is due
to a strongly bound exciton, and the second peak contains contributions
from higher excitons and from the onset of the continuum of inter-band
transitions (see below). The similarity between the RPA and GW+BS spectra 
stems exclusively from the strong broadening employed in the calculation.
The main peaks in the two spectra are at about the same position because 
of an almost-cancellation between the bandgap widening due to the 
GW-approximation and the red-shift of oscillator strength due
to excitonic effects. Comparison with Fig.~\ref{roughspecs} b) shows
that the shape of the GW+BS spectrum is in much better agreement
with experiment than the RPA spectrum. This underlines the importance
of excitonic effects in hBN. The influence of excitonic effects
becomes even more pronounced when we compare the real part of $\epsilon$.
Only the GW+BS calculation can reproduce qualitatively the shape of 
the experimental $\epsilon_1$.

While the shape of the experimental spectrum is well reproduced
by the LDA GW+BS calculation, the absolute position is not:
The main peak in the experiment is 0.65 eV higher
than in the LDA GW+BS spectrum. We suppose that the quasi-particle
gap is higher than the one predicted by the LDA GW method.
In order to check this hypothesis, we performed calculations based on 
EXX wave-functions. In Fig.~\ref{bandstruc}
we showed that the EXX GW gap is
wider by 0.45 eV than the LDA GW gap. This is manifested
in Fig.~\ref{roughspecs} c) where the
absorption spectrum having similar shape to the LDA GW+BS one
is blue-shifted by 0.45 eV \cite{ldascreening}.
The remaining difference of 0.2 eV with respect to the experimental
spectrum is most likely due to a short-coming of the
GW-approximation (possibly due to the neglect of vertex corrections).
In what follows we will present calculations based on 
LDA wavefunction and energies. Beyond the GW-correction,
we will introduce an additional scissor (0.65 eV) with
a value such that the first absorption peak at 6.1 eV, in agreement 
with experiments~\cite{zung76,Hoffman,tarrio,lauret} is properly
reproduced.

\begin{figure}
  \includegraphics*[width=.4\textwidth]{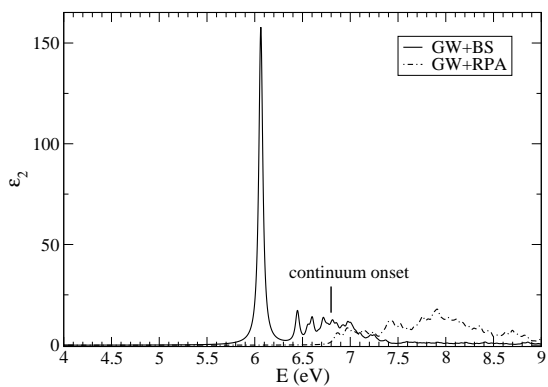}
  \caption{Absorption spectrum of hBN calculated with
a Lorentzian broadening of 0.025 eV. Solid line: LDA GW+BS calculation
(with additional scissor operator - see text). Dotted line: LDA GW RPA.} 
\label{finespec}
\end{figure}

In Fig.~\ref{finespec} we show the ``shifted" GW+BS
spectrum with a broadening of 0.025 eV. We also display
the GW RPA spectrum (corresponding to independent quasi-particle 
transitions). The onset of the latter at 6.8 eV marks the direct 
quasi-particle gap with respect to which the exciton binding energy 
is calculated. 
There is a sequence of bound excitonic peaks below the 
onset of the continuum. Most of the oscillator strength 
is collected by the first bound exciton which is doubly degenerate
and has a binding energy of 0.7 eV.
Upon broadening, the higher bound excitons and the absorption at the
continuum edge form together the second peak at 6.85 eV which has been
observed in experiments~\cite{Hoffman,tarrio} and can be seen
in Fig.~\ref{roughspecs}. Our results are therefore in good agreement
with the experimental data. It is also in good agreement with the
recent optical absorption measurement in ref.~\onlinecite{lauret},
assuming that their absorption peak at 6.1 eV is strongly broadened
due to sample quality of experimental resolution.
However, the interpretation of the experiments has changed:
While up to now, the spectrum of hBN has always been interpreted
in terms of a {\em continuum of inter-band transitions} with 
a {\em direct gap between 5.4 and 5.8 eV}, our calculations show
that the spectrum is dominated by {\em a strongly bound discrete
Frenkel excitons} with a {\em direct quasi-particle gap of 6.8 eV}.

\begin{figure}
  \includegraphics*[width=.3\textwidth]{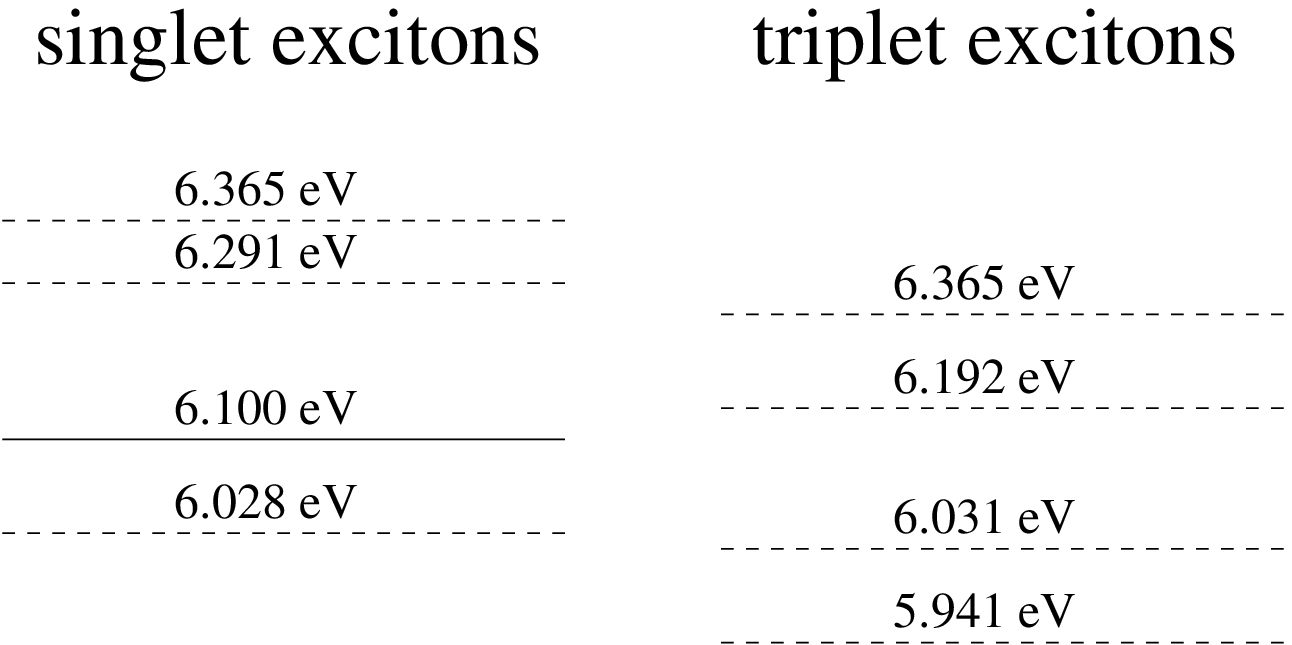}
  \caption{Energy scheme of optically active (solid line) and "dark" (dashed
lines) excitonic states around the first absorption peak at 6.1 eV.}
\label{scheme}
\end{figure}

Recently, Watanabe et al.~\cite{watanabe} presented optical absorption
data that is seemingly in contradiction with previous experimental data
and with our theoretical results. They find evidence for a direct
bandgap of 5.971 eV with a series of four bound excitons at 
5.822, 5.945, 5.962, and 5.968 eV. The binding energy of the first
exciton is therefore 0.149 eV in stark contrast with our value of 0.7 eV.
However, the absorption data of Ref.~\onlinecite{watanabe}
stops at 6 eV, making a comparison with previous absorption data difficult.
We suggest that the peaks that they observe are due to the excitation
of ``dark'' excitons, i.e., excitonic states that are forbidden by the
dipole selection rules but that might be accessible when the symmetry
of the crystal is broken through defects or limited sample quality
or when spin-orbit interaction is strong enough to lead
to spin-flip (i.e., excitation of triplet-excitons via intersystem crossing).
In Fig.~\ref{scheme}, we present schematically the levels of all
active and dark excitons in the energy range around the first
optically allowed exciton at 6.1 eV.
At 0.072 eV below this value a dark singlet exciton is found,
and at 0.159 and 0.069 eV below the first optically allowed singlet
exciton, two triplet excitons are found.
These energy values do not correspond exactly to the absorption
peaks of Ref.~\onlinecite{watanabe}, but they may explain qualitatively 
the occurrence of additional absorption peaks below 6.1 eV in the
work of ref.~\onlinecite{watanabe}. The limited accuracy of
the approximations used in the calculations together with the
possible role of 
exciton-phonon coupling could be explain the remaining discrepancies.
We note that our calculations are at variance with the calculations of
Ref.~\onlinecite{arnaud} that argue that the first absorption peak
is composed of 4 different optically active excitons.

\begin{figure}
  \includegraphics*[width=.4\textwidth]{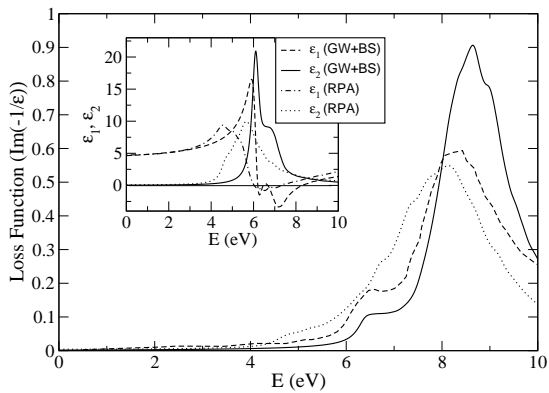}
\caption{Calculated loss function of hBN within the 
RPA (dotted line) and GW+BS (full-line).
Dashed line: experimental loss function
of ref.~\onlinecite{tarrio}. Inset: Real and imaginary part 
the dielectric function in the GW+BS and RPA methods.}
\label{lossfigure}
\end{figure}

Another important check of interpretation
is the direct comparison between calculated and
measured loss-function of hBN~\cite{tarrio}.  This is done
in Fig.~\ref{lossfigure} where we compare the RPA and GW+BS results.
The experimental loss function displays a peak around 8.4 eV which
is due to plasma-oscillations of the $\pi$-electrons. It has a shoulder
at 6.5 eV which has never been explained. A similar shoulder has been
recently observed in the electron-energy loss spectra of multi-wall
BN nanotubes~\cite{fuentes}. Inspecting the real part of the calculated
dielectric function, $\epsilon_1$, in the inset of Fig.~\ref{lossfigure}, 
the origin of this shoulder becomes clear. A peak in the loss function
occurs close to the energies where $\epsilon_1$ changes its sign from
negative to positive. While the RPA-calculation yields only one such
crossing and consequently only one plasmon peak, the excitonic 
effects lead to an additional
crossing at 6.5 eV which causes the shoulder of the $\pi$-plasmon peak,
in perfect agreement with experiments.

In conclusion, we have proposed a new scenario to explain the spectroscopic
properties of hBN. The main absorption peak of hBN
at 6.1 eV is due to a strongly bound exciton with a binding energy
of 0.7 eV. This means that the ``true'' minimum direct quasi-particle 
gap is 6.8 eV which is considerably larger than predicted 
previously \cite{blasebn}.  The additional ``excitonic'' peaks observed
in Ref. \onlinecite{watanabe} can be explained as due to a relaxation
of the selection-rules due to symmetry breaking and/or due to
spin-flip (singlet to triplet interconversion).
We also show that the electron-energy
loss spectra of hBN show a characteristic influence of excitonic
effects that manifest itself as a shoulder at lower energies of the main
$\pi$-plasmon peak.
We expect that these excitonic effects will also play an important 
role in the interpretation of experiments on EELS of isolated
BN-nanotubes \cite{arenal}.

The work was supported by the EU network of excellence 
NANOQUANTA (NMP4-CT-2004-500198) and the French GDR "nanotubes".
Calculations were performed at IDRIS (Project No. 51827) and CEPBA.
A.R. acknowledges the
Humboldt Foundation under the Bessel research award (2005),



\begin{thebibliography}{99}
\bibitem{gwarnaud}
There is a discrepancy with a recent
all-electron GW calculation~\cite{arnaud}
which reports a direct LDA GW-band gap
of 6.47 and an indirect gap of 5.95. The difference can possibly be
traced back to the fact that their in-plane static dielectric constant
is only 4.40 while we obtain 4.71 which is closer to the
experimental value of 4.95 eV \cite{geick}.
\bibitem{ldascreening}
As a consequence of the large EXX-band gap, the static dielectric
constant, $\epsilon_0=4.4$, calculated in EXX-RPA strongly 
underestimates the experimental value of 4.95 \cite{geick}. 
This leads to an overestimation of the excitonic
binding energy.  However, the resulting value of the BS+GW
calculation, $\epsilon_0=4.7$, is larger than the EXX-RPA
RPA value and close to the LDA-RPA value. In principle, 
one should perform BS-calculations with the updated 
screening until self-consistence is reached.
Since this is computationally too expensive, we decided to take
the LDA-RPA screening (which is sufficiently close to the
experimental one) in combination with the EXX wavefunctions
and energies.
\end{thebibliography}
\end{document}